# The FLOod Probability Interpolation Tool (FLOPIT): Improving Spatial Flood Probability Quantification and Communication Through Higher Resolution Mapping

**Mahkameh Zarekarizi[1*], K. Joel Roop-Eckart[2], Sanjib Sharma[1], and Klaus Keller[1,2]**
[1]Earth and Environmental Systems Institute, the Pennsylvania State University, University Park, PA, USA;
[2]Department of Geosciences, the Pennsylvania State University, University Park, PA, USA;
*Corresponding Author: mahkameh.zare@gmail.com, now at Jupiter Intelligence

**Abstract:** Understanding flood probabilities is essential to making sound decisions about flood-risk management. Many people rely on flood probability maps to inform decisions about purchasing flood insurance, buying or selling real-estate, flood-proofing a house, or managing floodplain development. Current flood probability maps typically use flood zones (for example the 1 in 100 or 1 in 500-year flood zones) to communicate flooding probabilities. However, this choice of communication format can miss important details and lead to biased risk assessments. Here we develop, test, and demonstrate the FLOod Probability Interpolation Tool (FLOPIT). FLOPIT interpolates flood probabilities between water surface elevation to produce continuous flood-probability maps. We show that FLOPIT can be relatively easily applied to existing datasets used to create flood zones. Using publicly available data from the Federal Emergency Management Agency (FEMA) flood risk databases as well as state and national datasets, we produce continuous flood-probability maps at three example locations in the United States: Houston (TX), Muncy (PA), and Selinsgrove (PA). We find that the discrete flood zones generally communicate substantially lower flood probabilities than the continuous estimates.

## 1. Introduction

Flooding drives sizable risks around the globe (Stromberg, 2007; IFRC, 2016). Between 1980-2013, global economic flood losses exceeded $ 1 trillion and resulted in approximately 220,0000 fatalities (Winsemius et al., 2015). Future flood risks are projected to increase driven by a complex interplay between changes in exposures, vulnerabilities, and hazards (Wing et al., 2018; Winsemius et al., 2015; Hallegatte et al., 2013, Winsemius et al., 2015; Hallegatte et al., 2013; Hirabayashi et al., 2013). Knowledge of flood risks can be a key driver in community participation in flood risk mitigation planning (Godschalk et al., 2003).

How one communicates flood probabilities can impact decision-making (Kjellgren, 2013; Zarekarizi, 2020). Flood probability maps are important sources of information about floods. The information communicated through these maps impacts decisions on where to build and whether to elevate structures to prevent flood damage and purchase flood insurance (FEMA, 2015; Zarekarizi, 2020). Flood maps typically consist of flood zones that bin continuous and spatially varying hazards into discrete flood zones (Ludy and Kondolf, 2012; Smith, 2000, Alfieri et al., 2013; Jafarzadegan et al., 2018; Woznicki et al., 2019). The outer edge of a zone is the maximum extent of a flood with a designated probability (i. e. the 1 in 100-year flood), while the inner edge of a zone is the maximum extent of a smaller magnitude, higher probability flood. The choice to communicate via flood zones can introduce a low-bias in flood probabilities (Ludy and Kondolf, 2012). For example, the Federal Emergency Management Agency (FEMA) in the United States, publishes 1 in 100-year (1% annual chance) and 1 in 500 years (0.2% annual chance) flood zones (FEMA, 2018). The maximum extent of the 1% annual chance flood designates the boundary between the 1% and 0.2% flood zones. Thus, the 1% annual chance flood zone is essentially the >1% annual chance flood zone, while the 0.2% annual chance flood zone is essentially the [1% , 0.2%] annual chance flood zone. Additionally, the "x" zones, also termed "area of minimal flood risk", have a <0.2% annual chance. This approach obscures the nature of continuous flooding probabilities with "in-out" flood zones (Ludy and Kondolf, 2012). The "in-out" approach can have serious consequences for flood hazard communication. Smith (2000) states that the 1 in 100-year flood zone "rapidly transformed in the community mind to a definition of flood free and flood-prone, with areas above the designated flood perceived to be flood-free—a misconception often reinforced by flood maps that shade only those portions that are subject to the designated flood". Adopting a flood zone approach neglects to communicate key information (e.g. Luke et al., 2018) and can communicate biased hazards (Ludy and Kondolf, 2012) (Figure 1).

The downward bias in the communicated flood probability and the binning associated with flood zones is well understood and communicated by flood mapping organizations (FEMA, 2018). However, downward biased and binned flood probabilities can present a communication barrier and the qualifiers communicated by flood mapping organizations may be ignored (Soden et al, 2017). This barrier can be particularly problematic if people ignore risks when they view the probability as falling below some threshold level of concern, as some research suggests (e.g., McClelland, Schulze, and Coursey,1993; Oberholzer-Gee,1998). One approach to reduce the downwards bias is to disaggregate flood zones into smaller zones. Wing et al. (2018) produce flood maps of the contiguous United States for 10 different probability floods, while FEMA maps between three and six different probability floods, but publishes only the 1 in 100 and 1 in 500-year flood maps. Increasing the number of probability zones decreases the downward bias and "in-out" issues associated with "binning" probabilities, but does not solve the underlying problem.

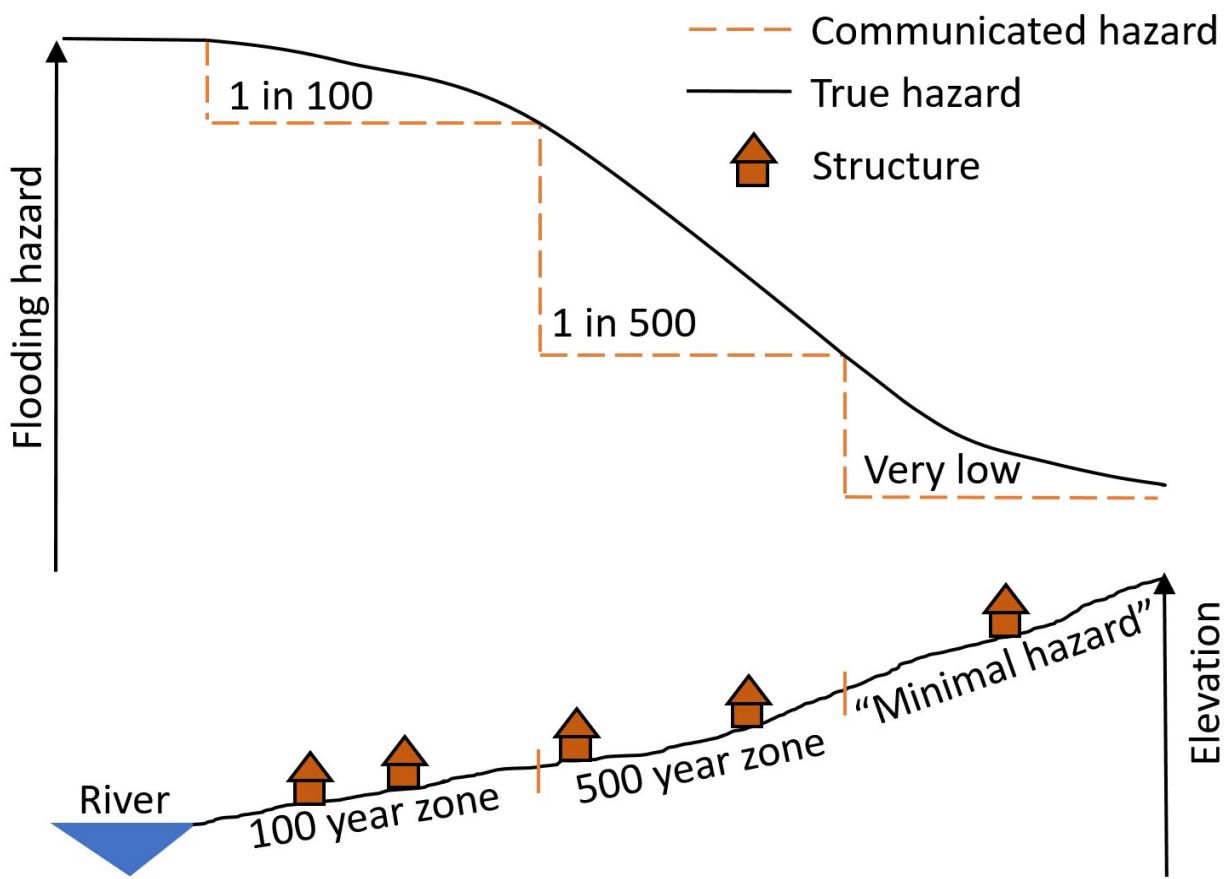

**Figure 1.** Schematic diagram of flood hazard communication via flood zone maps compared to the true hazard. The upper left y-axis represents the flood hazard. This is highest near the river and decreases as the elevation rises. The lower right y-axis represents elevation, in this simple illustration increasing with increasing distance from the river. Due to the way flood zones "bin" flood hazard into discrete zones, the communicated flood hazard is almost always lower than the true flood hazard.

FEMA is developing flood probability maps for their newest flood zones using a log-linear interpolation between flood probability and flood surface elevation (FEMA, 2015, 2018). However, methods of flood probability interpolation are not widely discussed or applied in the flood mapping literature (FEMA, 2018; Luke et al., 2017). In our experience, open source tools for flood probability mapping are difficult to find. Furthermore, while FEMA provides access to the flood probability maps (FEMA, 2015), these maps are not as easy to find as flood zones, nor are the flood probability maps typically used in homeowner decision-making.

Here we introduce, design, implement, and demonstrate the FLOod Probability Interpolation Tool (FLOPIT). FLOPIT is a simple tool to interpolate flood probabilities between flood zone boundaries to create continuous flood probability maps. We design FLOPIT to be hopefully useful in improving stakeholder flood probability communication and in research on decision-making. We demonstrate the feasibility and importance of flood probability interpolation by applying FLOPIT in three case study locations in the United States: a neighborhood on the Sims Bayous in Houston, Texas, the borough of Selinsgrove, Pennsylvania, and the borough of Muncy, Pennsylvania.

**2. Materials and Methods**

FLOPIT generates continuous flood probability maps from flood water surface elevation data associated with at least two return periods plus the digital elevation model (DEM) of the area of interest. Typically, FEMA provides flood surface elevation data for 10, 100 and 500-year flood events in the United States. Additionally, DEMs are widely available in varying accuracies across the United States and the world (Jafarzadegan et al., 2018; Yamazaki et al., 2019).

FLOPIT implements four main steps. Specifically, it:
(1) calculates water surface elevation by adding digital elevation data to user-inputted depth information,
(2) smooths water surface elevation data using an inverse weighted distance method,
(3) relates flood surface elevations associated with provided return periods for each raster cell to corresponding flood probabilities using the user's choice of monotonically increasing cubic spline interpolation (Fritch and Carlson, 1980) or log-linear interpolation (FEMA, 2018), and finally
(4) interpolates the flood probability for each cell from ground surface elevations derived from a DEM (Figure 2).

Errors or uncertainties in flood surface elevations or ground surface elevations can lead to flood probability map errors. In cases where the interpolated probability of a cell is outside the range of flood probability inputs, FLOPIT coerces a flood probability to the "zone" probability (the lowest possible probability bounded by the input data) if the cell is inside the flood extent. Cells that are outside the spatial extent of the flood maps are beyond the limit of extrapolation and coerced to a "not a number" (NA) value.

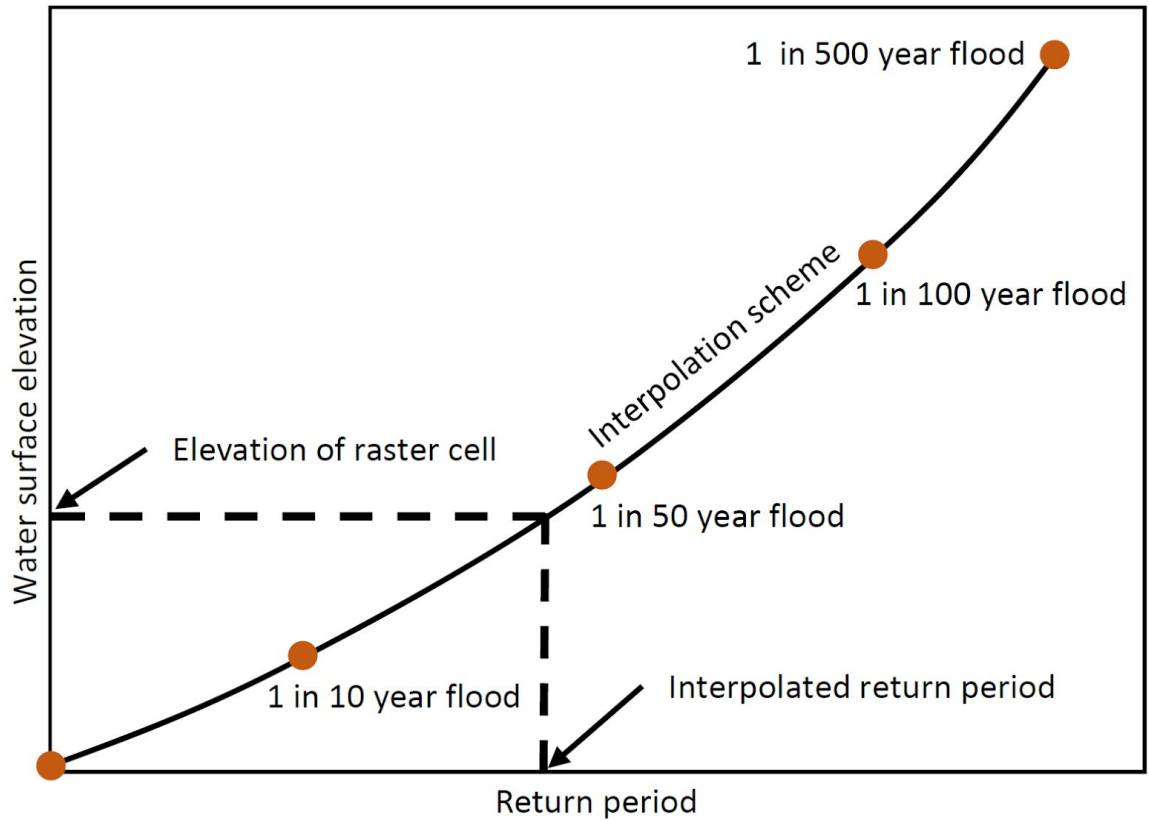

**Figure 2.** Conceptual diagram illustrating the FLOPIT flood surface elevation-probability interpolation approach. Flood surface elevation-probability data are not available for all flood probabilities. FLOPIT interpolates a water surface elevation to return period relationship between existing flood surface elevation-probability data. The continuous relationship is then used to interpolate flood probabilities for elevations between two flood surface elevations, producing a continuous flood probability map.

FLOPIT assigns a flood probability to all grid cells and outputs a raster file containing the flood probability map of the study area. The output flood probability map resolutions is limited by the resolutions of flood surface elevation data and the DEM. FLOPIT can handle a range of resolutions. The computation time scales roughly exponentially with increasing resolution. We use DEMs from the National Elevation Dataset (Gesch, 2002) and the Pennsylvania Spatial Data Access (PAMAP, 2020). We use flood surface elevations with known probabilities from the FEMA flood Map Service Center databases for each location.

## 3. Results

For the first case-study, we assess FLOPIT's performance at the Sims Bayou in Houston (TX). The Sims Bayou is one of a few areas where we could find published FEMA flood probability maps. FEMA uses log-linear interpolation (FEMA, 2018). We compare FLOPIT spline and log-linear interpolation results with FEMA flood probability maps at Sims Bayou. We also use FLOPIT to interpolate flood probability maps at the boroughs of Muncy and Selinsgrove (PA). These small Pennsylvania communities on the Susquehanna river face recurrent river flooding, but currently do not have FEMA-published flood probability maps, to the best of our knowledge.

For Sims Bayou, our study area is roughly one square kilometer. FLOPIT uses roughly 1 minute of wall time on a single core (I7-6700 4GHz CPU with 64 GB RAM) to interpolate the roughly 1 km$^2$ map, at 1/9 arc second (~10 ft or 3 m) resolution. The speed equates to roughly 1800 cells per second. Performance can vary considerably with degree of aggregation and number of cells. For this example, the information in the FLOPIT flood probability map diverges considerably from the information displayed in the FEMA flood zones (Figure 3).

We also generate FLOPIT interpolated flood probability maps of the boroughs of Muncy and Selinsgrove (PA), on the Susquehanna River (Figures A1 and A3). We notice substantial changes in communicated flood hazard both in Muncy and Selinsgrove. The FLOPIT interpolation for the roughly 37 and 11 km², and roughly 5 m and 2 m resolutions, for the boroughs of Muncy and Selinsgrove, took between 1 and 3 hours to run on a single core (I7-6700 4GHz CPU with 64 GB RAM), respectively. FLOPIT interpolated flood probabilities can be substantially greater than FEMA flood zone communicated probabilities for each of the three study sites (Figure 4, Figures A2 and A4).

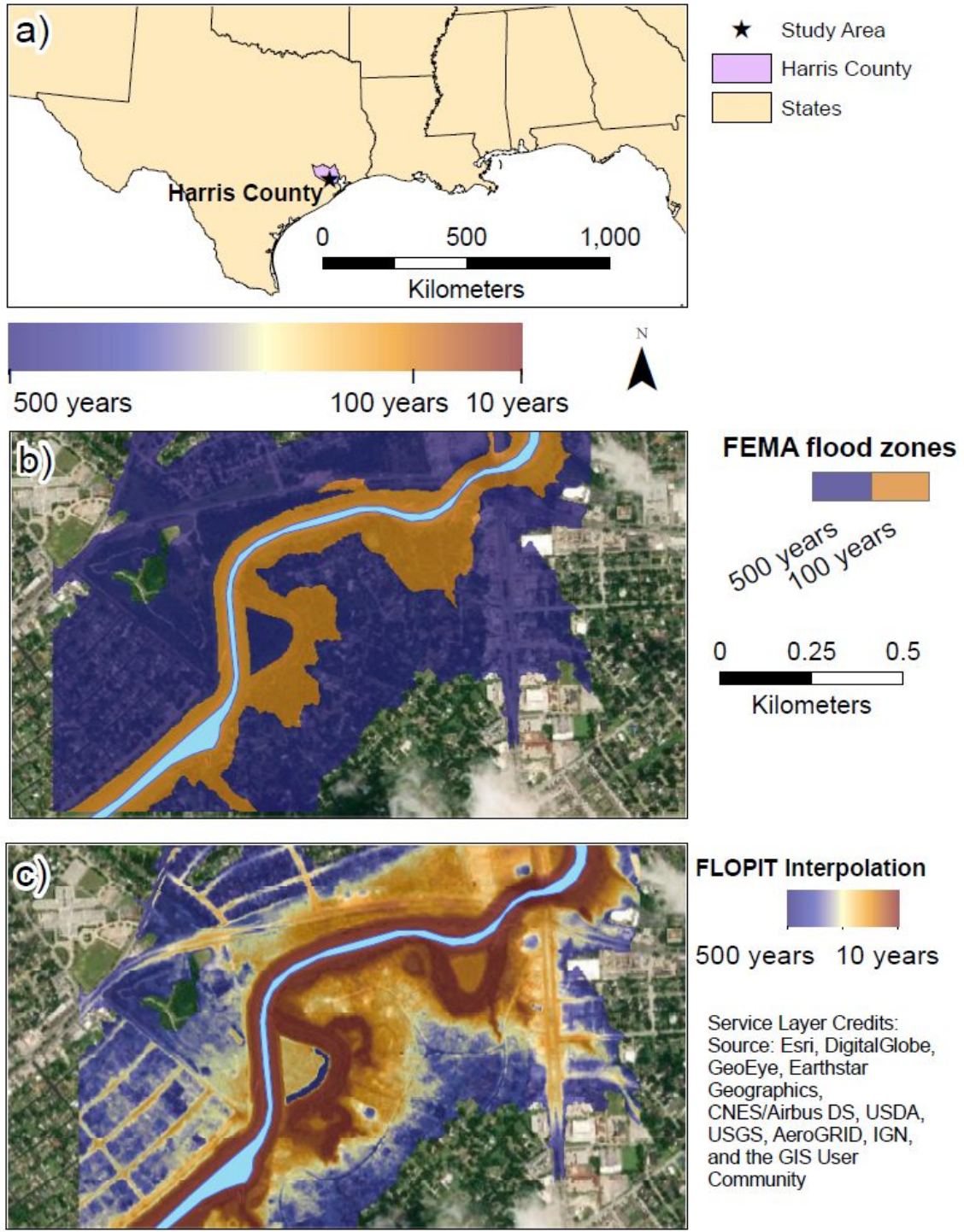

**Figure 3.** Map of a roughly 1.5 km reach of the Sims Bayou in Houston (TX). Panel a) shows the location of Harris County and the Sims Bayou. Panel b) shows the FEMA floodplains, derived from FEMA flood surface elevation data for the 1% and 0.2% annual chance (1 in 100-year and 1 in 500-year) floods. Panel c) shows the FLOPIT interpolated flood probabilities, from 10% to 0.2% annual chance (1 in 10-year to 1 in 500-year). Flood probabilities are always greater than or equal to the flood zone communicated probabilities.

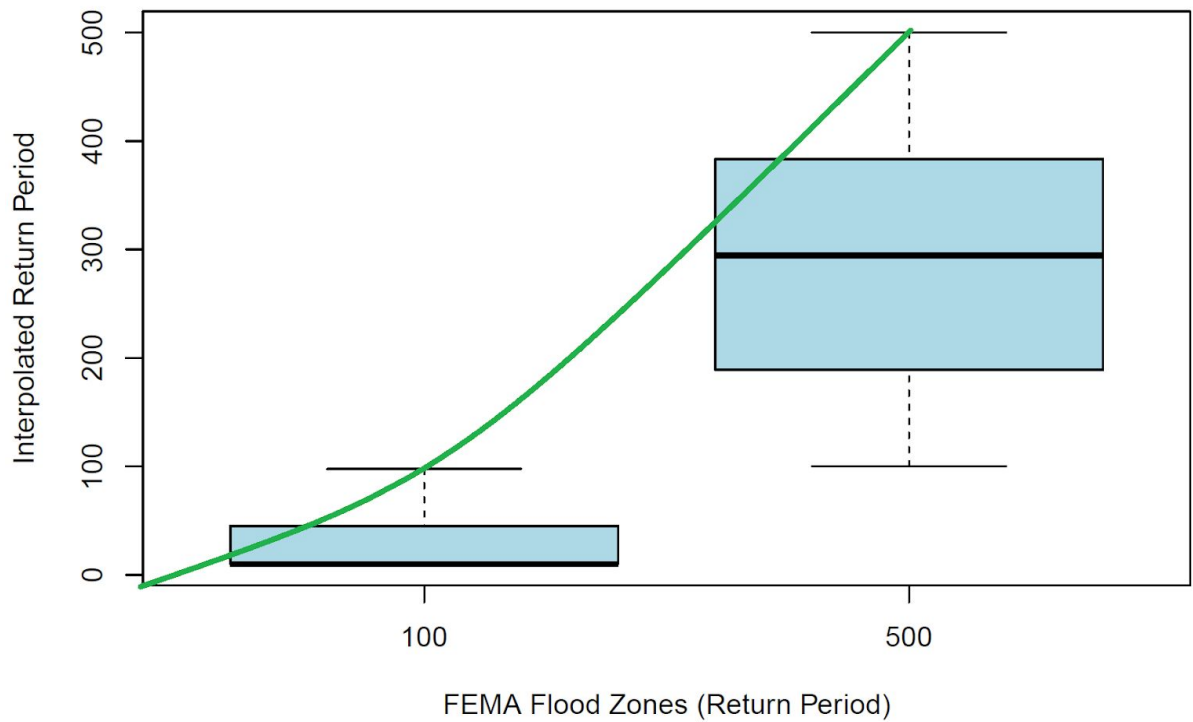

**Figure 4.** Box and whisker plot of the interpolated return period versus the FEMA flood zone return period for each Sims Bayou in Houston (TX), map pixel. Flood probabilities in the 1 in 100 (1% annual chance) flood zone range from 1 in 10 (10% annual chance) to 1 in 100 (1% annual chance), with an average flood probability of roughly 1 in 30 (3.3% annual chance). Flood probabilities in the 1 in 500 (0.2% annual chance) zone range from 1 in 100 (1%) to 1 in 500 (0.2%), and the average is roughly 1 in 300 (0.33% annual chance). The solid green line illustrates a hypothetical perfect relationship.

## 4. Discussion

Flood probability interpolation tools, such as FLOPIT, can create spatially resolved flood probability maps to help improve stakeholder communication and decision-making. FLOPIT provides spatially continuous flood hazard maps that compare favorably with flood zone maps while improving flood probability quantification. Continuous flood probability mapping also has the potential to improve flood hazard communication, stakeholder decision-making, and the setting of actuarial flood insurance rates.

FLOPIT's flexible resolution framework enables users to navigate the speed versus resolution trade-off with relative ease. The computation time scales roughly exponentially with increasing resolution. Decreasing the map resolution allows for faster flood probability mapping. This reduces the computational demands, at the cost of a degraded spatial resolution. One could conceive scenarios where flood probability data are needed at the home scale for individual flood risk assessments or the block scale for city-wide flood risk assessments. Future versions of FLOPIT will incorporate high performance computing approached for better scalability (Taheri et al., 2017; 2019)

Current flood hazard communication approaches typically rely on flood zones, assigning a single probability to an entire zone (FEMA, 2018). NFIP insurance rates are typically applied as a flat rate, such as the NFIP 500-year flood zone, or set by depth below a "base flood elevation" (the 1 in 100-year flood surface elevation). Flood probability interpolation can help to determine the probability of any single flood depth, or multiple flood depths. This allows for improved numerical integration over flood risk (Kron, 2005) for spatially resolved actuarial rate setting.

## 5. Conclusions

FLOPIT provides a flood probability interpolation tool that uses flood surface elevation-probability relationships and a digital elevation model to interpolate flood probabilities and produce flood probability maps. The tool provides a fast and flexible resource for producing continuous flood probability maps to aid flood hazard communication and quantification.

We demonstrate how flood zones can be a poor approximation to flood probability and can be downward biased in reported and communicated flooding probabilities. Flood probability interpolation tools can help address these issues by reducing the bias of spatially resolved flood probability maps. Refined flood probability maps can be useful to improve decisions, for example about where and how to build or whether to elevate a house, whether and how to change local zoning, and how to set fair flood insurance rates (Frazier et al., 2020, Karamouz et al, 2016, Kousky et al 2020, Thaler and Hartmann, 2016; Zarekarizi et al., 2020).

**Disclaimer:** All results, model codes, analysis codes, data, and model outputs used for analysis are freely available from https://github.com/pches/FLOPIT/tree/revision_2020 and are distributed under the GNU general public license. The datasets, software tools, results, and any other resources associated with FLOPIT and this manuscript are for academic research (not to inform decision-making) and provided as-is without warranty of any kind, express or implied. In no event shall the authors or copyright holders be liable for any claim, damages or other liability in connection with the use of these resources.

**Author Contributions:** Klaus Keller contributed to the study's conceptual framework, the interpretation of the results, and writing the manuscript. K. Joel Roop-Eckart contributed to the study's conceptual framework, wrote the code, performed the analysis, and contributed to writing the manuscript. Mahkameh Zarekarizi revised the code and edited the paper. Sanjib Sharma performed a code-review and edited the paper. All authors have read and agreed to the published version of the manuscript.

**Funding:** This study was co-supported by the US Department of Energy, Office of Science through the Program on Coupled Human and Earth Systems (PCHES) under DOE Cooperative Agreement No. DE-SC0016162 as well as the Penn State Center for Climate Risk Management. All errors and opinions (unless cited) are those of the authors and not of the funding entities.

**Acknowledgments:** We thank Caitlin Spence, Courtney Cooper, Jared Oyler, Irene Schaperdoth, Rob Nicholas, Karen Fisher-Vanden, and Skip Wishbone for valuable inputs.

**Conflicts of Interest:** The authors are not aware of conflicts of interest.

# Appendix A

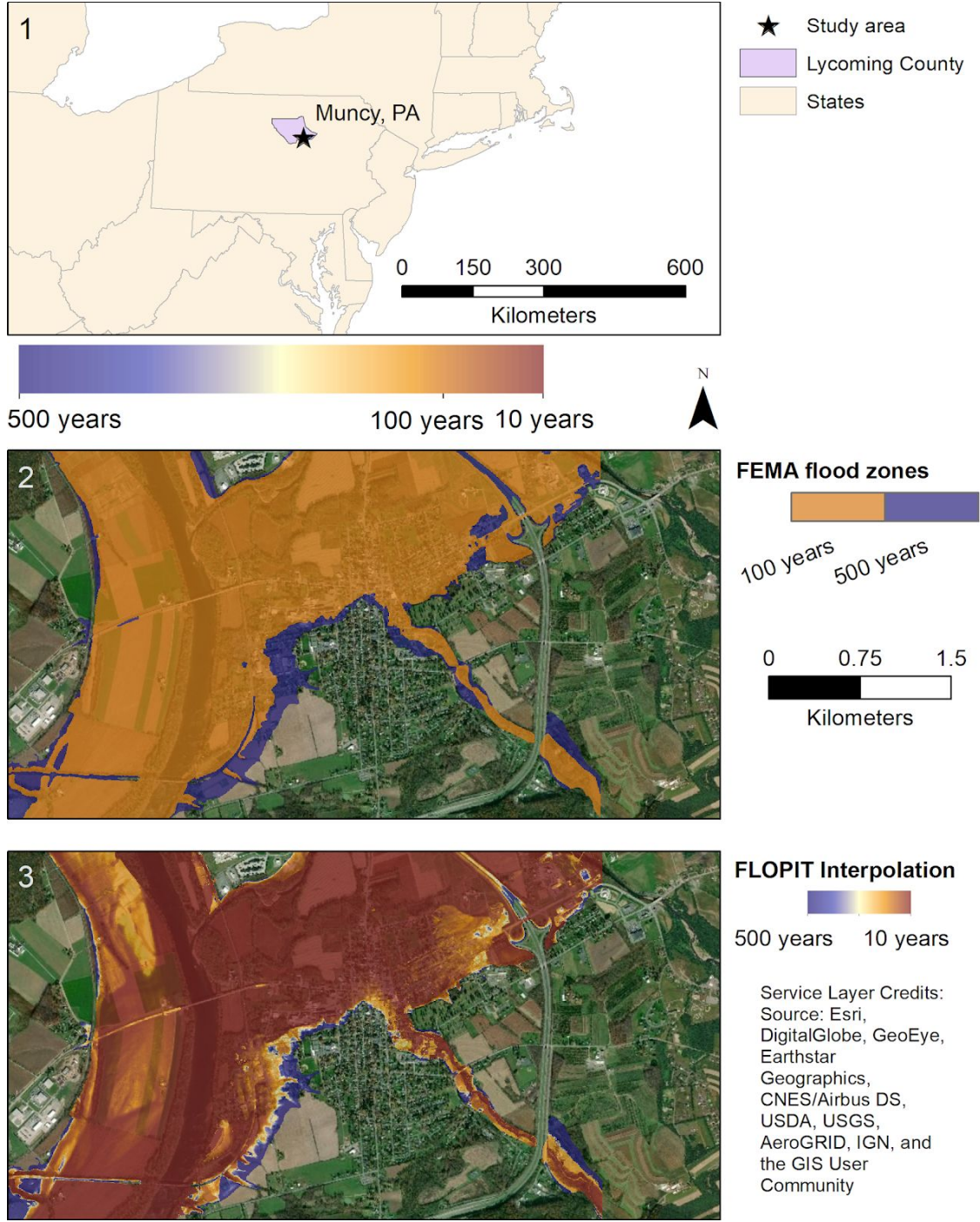

Figure A1. Map of a roughly 3 km reach of the Susquehanna River and tributaries at Muncy, PA. Panel 1 shows the location Lycoming County and Muncy, PA. Panel 2 shows the FEMA floodplains, derived from FEMA flood surface elevation data for the 1% and 0.2% annual chance (1 in 100-year and 1 in 500-year) floods. Panel 3 shows the FLOPIT interpolated flood probabilities, from 10% to 0.2% annual chance (1 in 10-year to 1 in 500-year). Flood probabilities are almost always higher than the flood zone communicated probabilities. The Borough of Muncy is located in the center of the map, with large northern sections inside the flood zones.

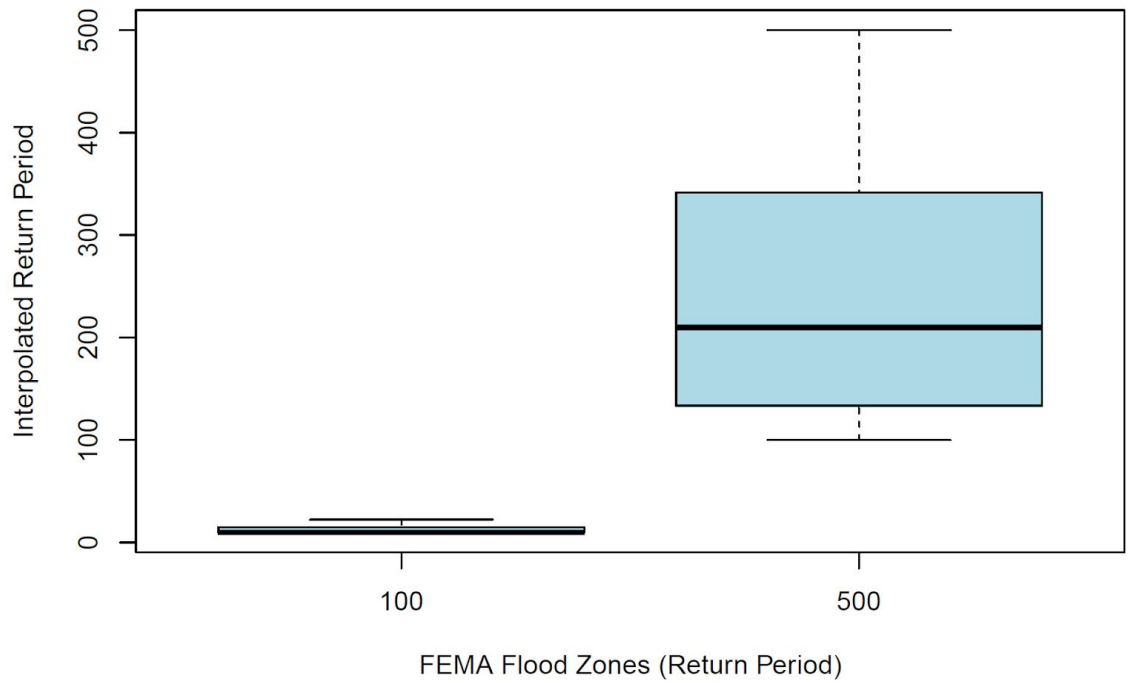

Figure A2. Box and whisker plot of the interpolated return period versus the FEMA flood zone return period for each map pixel of the Muncy map. Return periods in the 100 (1%) year flood zone range from 10 (10%) to 100 (1%), and the average return period is roughly 20 years (4%). Return periods in the 500 year zone range from 100 (1%) to 500 (0.2%), and the average is roughly 250 (0.4%) years.

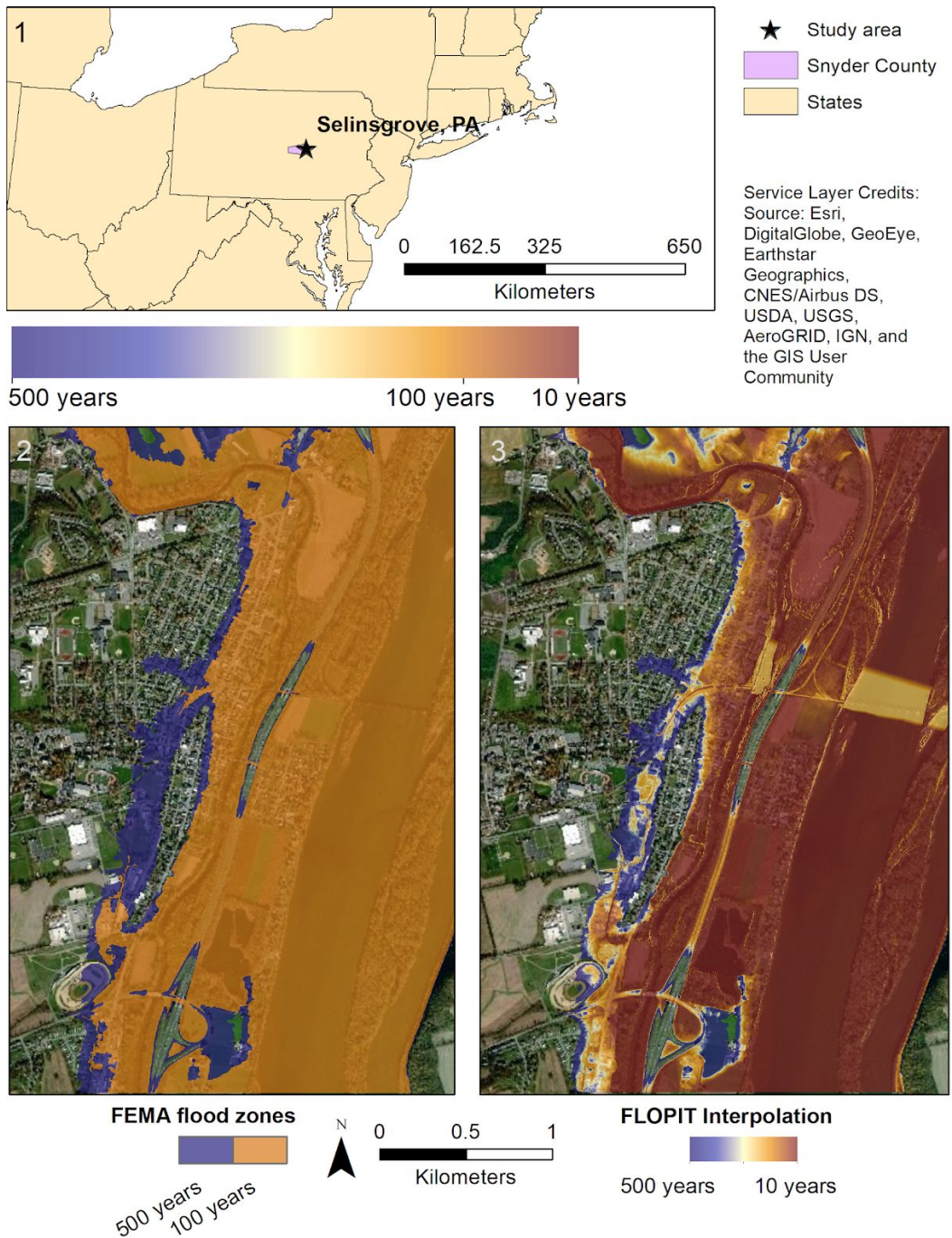

Figure A3. Map of a roughly 3 km reach of the Susquehanna River and tributaries at Selinsgrove, PA. Panel 1 shows the location Snyder County and Selinsgrove, PA. Panel 2 shows the FEMA floodplains, derived from FEMA flood surface elevation data for the 1% and 0.2% annual chance (1 in 100-year and 1 in 500-year) floods. Panel 3 shows the FLOPIT interpolated flood probabilities, from 10% to 0.2% annual chance (1 in 10-year to 1 in 500-year). Flood probabilities are almost always higher than the flood zone communicated probabilities. The Borough of Selinsgrove stretches from the north and west edges of the map to the river and just above the Selinsgrove Speedway, with large northern and eastern sections inside the flood zones.

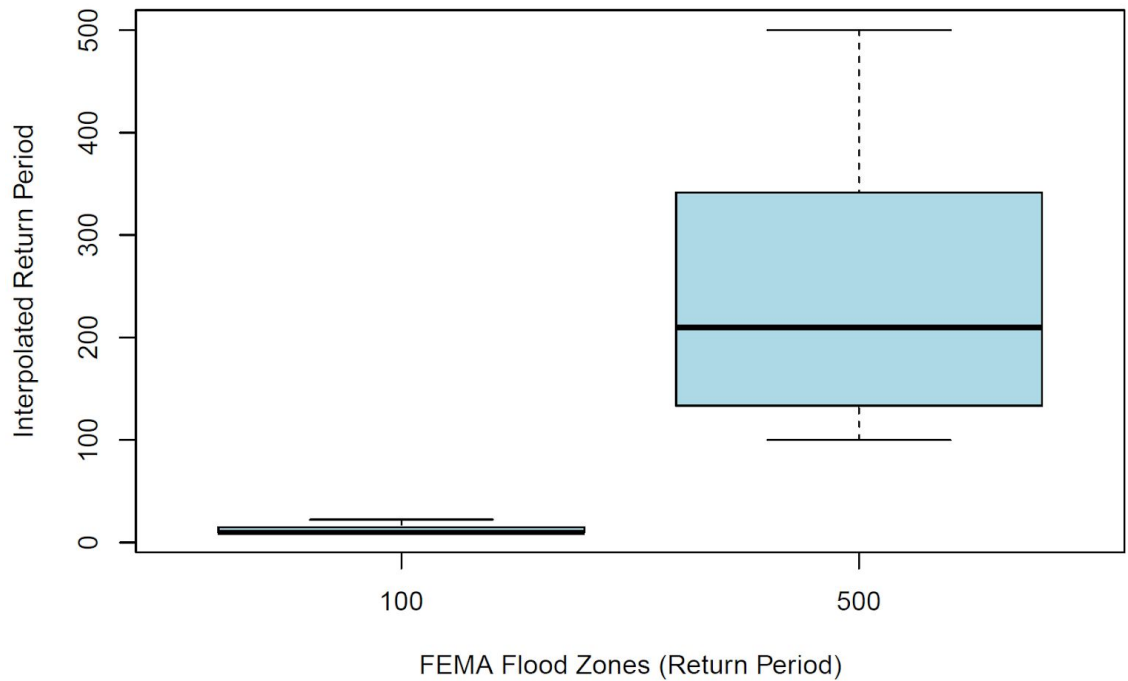

Figure A4. Box and whisker plot of the interpolated return period versus the FEMA flood zone return period for each map pixel of the Selinsgrove map. Return periods in the 100 (1%) year flood zone range from 10 (10%) to 100 (1%), and the average return period is roughly 20 years (4%). Return periods in the 500 year zone range from 100 (1%) to 500 (0.2%), and the average is roughly 250 (0.4%) years.